\documentclass{article}
\usepackage[T1]{fontenc} 
\usepackage[utf8]{inputenc} 
\usepackage{ismir,amsmath,cite,url}
\usepackage{graphicx}
\usepackage{color}
\usepackage{booktabs}
\usepackage{amssymb}

\usepackage{pgfplots}

\usepackage{xcolor} 

\usepackage{hhline}
\usepackage{multirow}

\usepackage{float}

\usepackage{enumitem }

\pgfplotsset{compat=1.17}
\usepackage{tikz}
\usetikzlibrary{pgfplots.groupplots}
\usepgfplotslibrary{groupplots}
\pgfplotsset{compat=newest} 
\usetikzlibrary{matrix} 

\usepackage{lineno}

\usepackage{comment}

\title{{Diff-A-Riff}: Musical Accompaniment Co-creation via Latent Diffusion Models}

\multauthor
{Javier Nistal$^1$ \hspace{1cm} Marco Pasini$^2$ \hspace{1cm} Cyran Aouameur$^1$} { \bfseries{Maarten Grachten$^1$ \hspace{1cm} Stefan Lattner$^1$}\\
$^1$Sony Computer Science Laboratories, Paris, France\\
$^2$Queen Mary University of London, UK\\
{\tt\small javier.nistal@sony.com}
}

\def\authorname{J. Nistal, M. Pasini, C. Aouameur, M. Grachten, and S. Lattner}

\usepackage[bookmarks=false,pdfauthor={\authorname},pdfsubject={\papersubject},hidelinks]{hyperref}

\begin{document}
\maketitle
\begin{abstract}
Recent advancements in deep generative models present new opportunities for music production but also pose challenges, such as high computational demands and limited audio quality. Moreover, current systems frequently rely solely on text input and typically focus on producing complete musical pieces, which is incompatible with existing workflows in music production. To address these issues, we introduce Diff-A-Riff, a Latent Diffusion Model designed to generate high-quality instrumental accompaniments adaptable to any musical context. This model offers control through either audio references, text prompts, or both, and produces 48kHz pseudo-stereo audio while significantly reducing inference time and memory usage. We demonstrate the model's capabilities through objective metrics and subjective listening tests, with extensive examples available on the accompanying website.\footnote{\url{sonycslparis.github.io/diffariff-companion/}}
\end{abstract}

\section{Introduction}\label{sec:introduction}


Deep generative modeling has recently made significant strides, greatly expanding the toolbox for synthesizing visual and auditory art \cite{makeavideo, dalle3, musiclm, musicgen, stableaudio, audioldm2} and signaling a new era of enhanced creative expression.
These technologies promise more intuitive, high-level control over digital creations, yet their deployment in music production comes with inherent challenges. 
Generative music systems frequently rely solely on text inputs for control and typically focus on generating complete musical pieces rather than individual sounds or instruments. This approach can limit their integration into existing musical workflows and may compromise the artist's control over the final product. 
Furthermore, the computational demands of these advanced models often necessitate access to specialized hardware or online services. Additionally, they often fail to meet professional audio standards, such as true stereo output at 48 kHz.


In this paper, we introduce Diff-A-Riff, a novel Latent Diffusion Model designed for generating single-instrument accompaniments. A distinct feature of our approach is the ability to condition on musical audio contexts. This specific form of control crucially allows the music to dynamically adapt to the artist's style, enabling a more personalized creation process. Additionally, the model supports conditioning using joint text-and-audio embeddings from CLAP \cite{clap}, which can be derived from either textual descriptions or audio references, providing versatile input options for directing the music generation.


At the core of Diff-A-Riff are two pivotal technological elements. First, the efficiency of a Consistency Autoencoder with a high compression rate enhances the system's performance in terms of inference time and memory usage \cite{m2l2}. Second, the model employs the expressiveness of Elucidated Diffusion Models (EDMs), known for their robust handling of complex data distributions and improved efficiency in model parameterization and inference \cite{edm}.



We validate Diff-A-Riff through comprehensive evaluations, assessing its performance in ablation studies using objective metrics, and we compare it to other models and estimate context adherence using subjective listening tests. The results, detailed in Sections \ref{sec:results} and \ref{sec:conclusion}, demonstrate that our model not only achieves state-of-the-art audio quality (statistically not distinguishable from real audio) but also effectively adapts to various conditional settings confirming its potential for practical applications in music production.

The paper is organized as follows: after a review of related work in Section \ref{sec:sota} and background in Section \ref{sec:background}, we describe our methodology in Section \ref{sec:method}. We then present our results in Section \ref{sec:results} and conclude with a discussion and potential future research directions in Section \ref{sec:conclusion}.

\section{Related Work}\label{sec:sota}

\noindent\textbf{End-to-end models}. The landscape of generative models for music has undergone transformative advancements in recent years. End-to-end Autoregressive Models (AMs) have traditionally been at the forefront of sound fidelity, diversity, and long-term coherence \cite{wavenet, samplernn}. Nonetheless, their high computational demands render AMs unsuitable for music production settings (i.e., sample rate $\geq$ 44.1\,kHz, stereo). In contrast, Generative Adversarial Networks \cite{gan}
and Variational Autoencoders \cite{vae}
 exhibit exceptional generation speed at high sampling rates \cite{rave, musika, drumgan}, positioning them as valuable assets for commercial music production technologies \cite{rave, drumganvst}. However, these strategies typically require simple datasets with reduced diversity \cite{rave, musika}, and often restrict generation to fixed lengths \cite{drumgan, drumganvst}. Recently, diffusion models showed a balanced equilibrium between generation quality, diversity, and efficiency \cite{crash, noise2music, edmsound, archiSound}. Nevertheless, these rely on an iterative denoising process that, while faster than AMs, still demands long and heavy computations.
 
 \vspace{0.1cm}
\noindent \textbf{Latent models}. To address the challenges inherent in end-to-end modeling, generative models have recently pivoted towards operating on compressed representation spaces learned via autoencoders \cite{musiclm, musicgen, stableaudio, audioldm2, jukebox}. By doing so, generative systems can allocate representational capacity separately for learning immediate auditory characteristics of sound and longer-term music structure. Additionally, they facilitate the interpretation and integration of multi-modal control data, such as text \cite{musiclm, stableaudio, audioldm2}, audio \cite{stemgen, bassnet2}, or melody \cite{musicgen}. Within this evolved framework, AMs leverage discrete representation spaces crafted through vector-quantized variational autoencoders 
 \cite{soundstream, encodec}, resulting in faster models with better long-term structure~\cite{musiclm, musicgen}. 
Recent developments have equipped AMs with parallel decoding using masked token modeling techniques \cite{vampnet, stemgen, magnet}, enabling sample rates as high as 44.1\,kHz with acceptable inference speed. 

Latent Diffusion Models (LDMs) also operate on compressed representation spaces, which are typically continuous \cite{stableaudio, audioldm2, mousai}. This evolution has catalyzed the development of various LDMs capable of generating high-resolution musical audio with long-term structure \cite{audioldm2, mousai, jen1, stableaudio}. Notably, some works can generate audio at sampling rates as high as 48\,kHz \cite{audioldm2} and stereo \cite{mousai, jen1}. Other works like Stable Audio \cite{stableaudio} improve inference efficiency, enabling the generation of 44.1\,kHz sampling rate and stereo audio at an unprecedented speed.\footnote{95 seconds of audio in 8 seconds on an A100 GPU} 
Following this spirit, our system leverages a pre-trained Consistency Autoencoder\cite{m2l2} which enables Diff-A-Riff to function within a highly compressed representation space, allowing faster generation than previous systems. Further, our LDM employs the framework of Elucidated Diffusion Models (EDMs) \cite{edm, edmsound}, a departure from the Denoising Diffusion Implicit Models (DDIMs) \cite{ddim} used in previous approaches \cite{audioldm2, stableaudio, mousai}.

 \vspace{0.1cm}
\noindent \textbf{Control mechanisms}. As evidenced by the state-of-the-art, text prompts currently serve as the most common interface for users to guide audio generative models \cite{musicgen, musiclm, stableaudio, audioldm2}. To facilitate finer control, Jukedrummer \cite{Jukedrummer} and Music ControlNet \cite{mcontrolnet}
utilize time-varying controls such as rhythmic and dynamic envelopes and melodic lines. By exploiting the semantic properties of multi-modal text-and-audio spaces, recent works propose zero-shot solutions to music editing via latent space manipulations \cite{musicmagnus} and inversion methods \cite{manor}. Alternative approaches to control pretrained models include inference-time optimization \cite{ditto} or guidance \cite{cmp}. Another method for influencing audio output involves conditioning on audio signals, a technique primarily used in style transfer and accompaniment tasks. In style transfer, the objective is to emulate specific aspects of the source audio, e.g., melody \cite{musicgen}, timbre \cite{bassnet2}. For accompaniment, the focus is on generating musical content that complements or enhances the conditioning audio \cite{bassnet, bassnet2, singsong, stemgen, DrumNet}. Recent works attempting joint music generation and source separation also exhibit compositional capabilities such as accompaniment generation without requiring paired data \cite{postolache, multisourceddm}.
Inspired by these control mechanisms, our system introduces conditioning on audio and textual features derived from CLAP \cite{clap} alongside audio signals that serve as music context, widening the scope of generative capabilities, e.g., accompaniment generation, text-driven generation, and style transfer.
\begin{figure*}[t]
\vspace{-0.3cm}
\centering
\begin{minipage}{\textwidth}
  \centering
  \includegraphics[width=\linewidth]{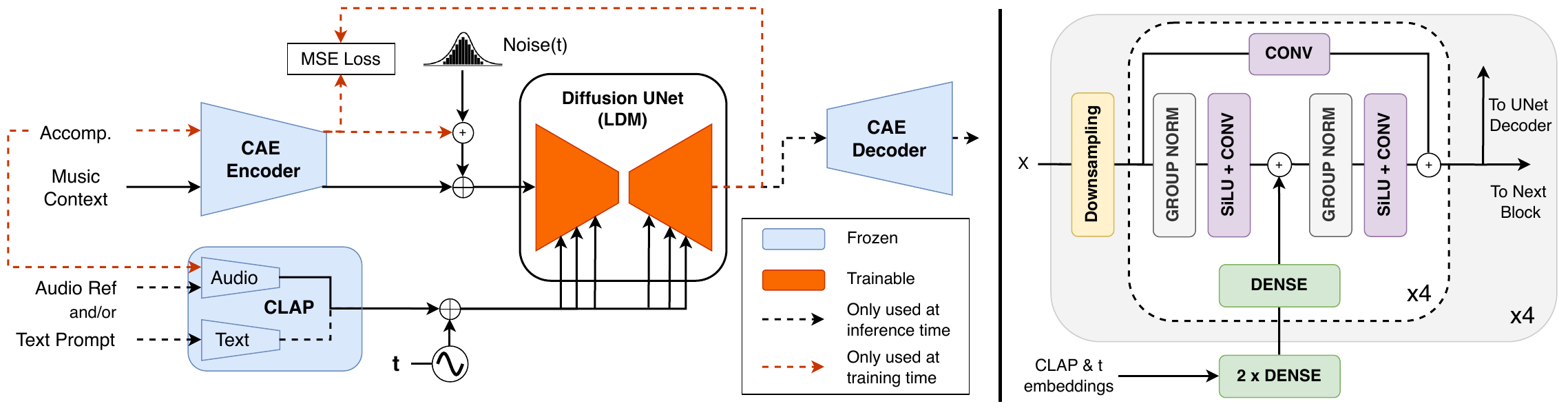}
  
  \label{fig:diffariff-a}
\end{minipage}%
\caption{Overview of Diff-A-Riff. \textbf{Left}: The CAE Encoder transforms the music context into a compressed representation, concatenated with a noisy sample, and further processed through a multi-scale U-Net. At each scale, conditional CLAP and time-step embeddings are integrated through a feature-wise linear transformation. The generated latent sequence is decoded via the CAE Decoder. We highlight frozen components in blue and trainable elements in orange. Text prompting is only used at inference. \textbf{Right}: The encoder architecture comprises four down-sampling blocks with four convolutional and group norm layers with skip connections. The decoder mirrors this architecture.}
\label{fig:diffariff}
\vspace{-0.3cm}
\end{figure*}



\section{Background}
\label{sec:background}

In this section, we provide a brief overview of Consistency Models and Denoising Diffusion Models. For an in-depth explanation, we encourage the reader to review the corresponding references.

 \vspace{0.1cm}
\noindent \textbf{Consistency Models (CMs)} \cite{consistency_models, improved_consistency_models} are a novel class of generative models that can produce high-quality samples in a single forward pass without adversarial training. CMs learn a mapping between noisy and clean data samples via a probability flow Ordinary Differential Equation (ODE)~\cite{ddim}. Given a noise level $t$, the consistency function $f$ transforms a noisy sample $x_t \sim p_t(x)$ to a clean sample $x \sim  p_{data}(x)$ by mapping $f(x_t, t) \mapsto x$. This consistency function is approximated by a neural network $f_\theta(x_t, t)$ with parameters $\theta$. It must satisfy the boundary condition $f_\theta(x, t_{\text{min}}) = x$ and is trained by minimizing the discrepancy between its output and a teacher CM at adjacent noise levels $t_i$ and $t_{i+1}$.

 \vspace{0.1cm}
\noindent \textbf{Denoising Diffusion Models (DDMs)} \cite{ddpm} are generative models originally inspired by the concept of thermodynamic diffusion \cite{diffusion_original}. DDMs first add noise to data in a \textit{forward} diffusion process and then use a neural network to \textit{reverse} this process by removing the noise iteratively. The forward diffusion process is detailed by a Stochastic Differential Equation (SDE), introducing noise to the original data $x_0$ over $T$ steps, resulting in a noisy version $x_T$. This process is defined by $dx_t = f(x_t, t)dt + g(t)dB_t$. Here, $dB_t$ is the increment of a Wiener process (the random noise), $f(x_t, t)$ is the drift term, $g(t)$ is the diffusion term, and $t$ represents the diffusion time step. The reverse process aims to reconstruct the original data from its noisy version by removing the noise. This is achieved by modeling the score of the data distribution, i.e., the gradient of the log probability density function of the noisy data with respect to the data itself, $\nabla_x \log p(x|t)$. The reverse process is described by another SDE, which guides the denoising $dx_t = [f(x_t, t) - g(t)^2 \nabla_x \log p(x_t|t)]dt + g(t)dB_t$, where a neural network $g_{\theta}$, with parameters $\theta$, is trained to estimate this score function, i.e., $g_{\theta}(x_t, t) \approx \nabla_{x} \log p(x_t|t)$. During inference, by performing this process iteratively, we can progressively transform pure noise inputs into data points following the training data distribution.
    
\section{Methodology}
\label{sec:method}

\subsection{Dataset}
We train our model on a proprietary dataset comprising 12,000 multi-track recordings of diverse music genres (e.g., pop/rock, R\&B, rap, country). Each multi-track has various instrument tracks, including bass, guitars, pianos, vocals, and more. We resample each track to 48\,kHz, convert it to mono and segment it into overlapping windows of approximately 10 seconds with a 3-second hop size.
In training, we randomly select a target \textit{accompaniment} track (excluding vocals) and construct the music \textit{context} by mixing a random subset of the remaining tracks for each segment. We apply this data segmentation and sampling strategy offline to obtain 1M training pairs of audio segments. Following the same methodology, we derive a validation set from 1,200 multi-tracks. 

\subsection{Diff-A-Riff}

\subsubsection{Consistency Autoencoder}
\label{sec:cae}
In this work, we employ a consistency model-based Autoencoder (CAE). We use it pre-trained and freeze its parameters to train a generative model on its latent embeddings (see Fig.~\ref{fig:diffariff}). The  CAE encodes audio samples into a continuous representation space with a $64\times$ compression ratio.
It operates on complex Short-Time Fourier Transform (STFT) spectrograms, with real and imaginary components as separate channels. The architecture uses convolutional residual blocks interleaved with down/up-sampling layers. The \textit{CAE Encoder} produces 64-dimensional encodings in the range $(-1, 1)$ with a sample rate of 12\,Hz for 48\,kHz input audio.
The model has $\sim58$ million parameters and is trained following the consistency training framework \cite{improved_consistency_models}. For a detailed description of the architecture and training procedure, we refer the reader to the original reference \cite{m2l2}.

\subsubsection{Latent Diffusion Model}
We train a Latent Diffusion Model (LDM) on the latent space learned by the CAE. The proposed LDM follows the framework of Elucidated Diffusion Models (EDMs) \cite{edm}, a departure from DDIMs~\cite{ddim} for improved model parametrization and inference.
The architecture follows DDPM++ \cite{NCSN++}, an upgraded version of the originally proposed Diffusion Probabilistic Model \cite{ddpm}. We only adapt the network's input dimensionality to that of the CAE's latent space. The input to the model is a channel-wise concatenation of the CAE-compressed music context and noisy input ($64\times2$ channels, see Sec.~\ref{sec:cae}). Also, we increase the dimensionality of the conditional embedding input with that of CLAP \cite{clap} (i.e., 512 channels). 
Our UNet is composed of four down/up-sampling blocks with convolutional layers and skip connections, both for the encoder and the decoder (see Fig.~\ref{fig:diffariff} Right). Self-attention is employed in the penultimate resolution layer. We use 512 base channels and double their number at each resolution block. Additionally, the model relies on two dense layers to project the concatenation of CLAP embeddings and the sinusoidal denoising step embeddings into a joint representation. The resulting embedding is used in all down/up-sampling blocks to condition the denoising process as illustrated in Fig~\ref{fig:diffariff}.

\subsubsection{Training}
Fig.~\ref{fig:diffariff} provides a high-level overview of Diff-A-Riff's setup. Given a pair of input \emph{context} and target \emph{accompaniment} audio segments, the model is trained to reconstruct the \emph{accompaniment} given the \emph{context} and a CLAP embedding derived from a randomly selected sub-segment of the target itself. This prevents the model from relying on CLAP for temporal alignment. In order to use Classifier-Free Guidance (see Sec.~\ref{sec:cond-inference}) and allow the model to optionally operate unconditionally, we drop the audio context and clap embeddings both with a 50\% probability. We train Diff-A-Riff over 1M iterations using a batch size of 256 (2 weeks on a single RTX 3090 GPU). We use AdamW \cite{adamw} as the optimizer and a base learning rate of $10^{-4}$. We use a learning rate schedule with an initial warm-up phase and a reduce-in-plateau process that decreases the learning rate to a minimum value of $10^{-6}$. We keep an Exponential Moving Average (EMA) on the weights with a momentum of $0.9999$. The resulting model has ~500M parameters (including the CAE and not CLAP) and occupies 3\,GB of memory.

\subsection{Evaluation}
Objective comparison of Diff-A-Riff with existing state-of-the-art models \cite{stableaudio, audioldm2, musicgen} is challenging as these are generally trained to perform a substantially different task (generation of fully mixed music with no accompaniment conditioning). Even though StemGen \cite{stemgen} and SingSong \cite{singsong} are trained to generate accompaniments, their implementation and pretrained weights are not publicly available.
Therefore, we focus on subjective listening tests to compare with available music generation models. Additionally, we perform objective evaluations to analyze different inference parameters (e.g., number of diffusion steps, conditioning information; see Section \ref{sec:cond-inference}) to understand which configurations of our proposed model perform the best on our set of metrics. We then apply the gained insights in order to generate the samples that are proposed in the user studies. In the following sections, we describe how we generate the samples used for evaluation, the objective metrics, the listening test methodology, and the baselines we compare against.

\subsubsection{Inference Configurations}
\label{sec:cond-inference}
In this section, we describe the inference configurations that we use for the objective and subjective evaluations.

 \vspace{0.2cm}
\noindent \textbf{Conditioning Signals}: Different conditioning signals are evaluated. $\textit{CLAP}_A$ refers to the audio-derived CLAP embeddings, which are obtained by using CLAP to encode real audio from a track of the evaluation set.
We can also condition the model on text-derived CLAP embeddings, despite them never being fed during training to the model, since CLAP offers a joint embedding space for both modalities. Because our dataset does not contain audio/text pairs, we create $\textit{CLAP}_T$ embeddings by asking ChatGPT to write text descriptions of single-stem tracks. Finally, \textit{Context} refers to the conditioning signal obtained by solely encoding the music context into the CAE Encoder. 

 \vspace{0.2cm}
\noindent \textbf{Classifier-Free Guidance (CFG)} \cite{cfg} allows to improve generation quality by increasing the influence of conditioning signals in the sampling process. 
Given the guidance strength CFG, we implement guidance as $x_{t-1} = f_{\theta}(x_t) + \text{CFG}\cdot(f_{\theta}(x_t, c) - f_{\theta}(x_t))$. At inference time, we can use different guidance strengths for \textit{Context} and \textit{CLAP} embeddings, denoted as $\text{CFG}_\textit{Context}$ and $\text{CFG}_\textit{CLAP}$, respectively.

 \vspace{0.2cm}
\noindent \textbf{Number of Diffusion Steps}: At inference time, the number of denoising steps $T$ allows to trade between audio quality and generation speed.


 \vspace{0.2cm}
\noindent \textbf{Pseudo-Stereo Generation}: We generate pseudo-stereo audio by denoising until a given diffusion time step, and then by independently concluding a stochastic denoising process twice, one for each audio channel. We define the \textit{stereo width} as the proportion of denoising steps used for stereo generation over the total number of steps. In the user study, we set this parameter to $0.4$.

\subsubsection{Objective Metrics}
We evaluate Diff-A-Riff through objective metrics to assess various aspects of the generated audio. These include the standard \textit{Squared Maximum Mean Discrepancy} (MMD2)~\cite{kid} and \textit{Fréchet Audio Distance} (FAD)~\cite{fad} for audio quality as well as \textit{Density} and \textit{Coverage}~\cite{den_cov} for evaluating fidelity and diversity. To study the system's responsiveness to text prompts, we employ the \textit{Clap Score} (CS)~\cite{Make-An-Audio}, which calculates the cosine similarity between text and audio embeddings. In order to evaluate the alignment of the generated accompaniment with the context, we employ the \textit{Audio Prompt Adherence} (APA) \cite{apa}, a 
metric for evaluating accompaniment systems. 
All metrics are calculated by averaging five batches of 500 candidate samples. We use CLAP \cite{clap} as the embedding space for metrics that compare distributions (like MMD2 and FAD) using a reference set of 5,000 real audio examples.

\begin{table*}[t]
\begin{small}
 \begin{center}
 \begin{tabular}{cccccccc}
  \toprule
   & Cond. Signal & $\downarrow$ MMD2$^a$ & $\downarrow$ FAD & $\uparrow$ \textit{Coverage} & $\uparrow$ \textit{Density} & $\uparrow$ APA & $\uparrow$ CS\\
  \midrule

    {\textit{Real}}       
                & Original acc. & 0.00 & 0.02 & 0.18 & 1.03 & 0.93 & 1.00 \\
    \midrule
    \textit{Lower bound}       & - & 64.32$^b$ & 1.67$^b$ & 0.00$^b$ & 0.00$^b$ & 0.11$^c$ & -0.07$^c$ \\
    \midrule
    \midrule
    \multirow{5}{*}{\textit{Diff-A-Riff} $_{T=30}^{\text{CFG}=1.25}$} 
        & $\textit{CLAP}_{A} + \textit{Context}$    & \textbf{0.22} & \textbf{0.03} & \textbf{0.17} & \textbf{1.00} & \textbf{0.92} & - \\
        & $\textit{CLAP}_{T} + \textit{Context}$     & 3.96 & 0.17 & 0.05 & 0.32 & 0.20 & \textbf{0.25} \\
        & \textit{Context} only    & 4.87 & 0.24 & 0.05 & 0.37 & 0.23 & - \\
        & $\textit{CLAP}_{A}$ only       & 0.36 & 0.03 & 0.14 & 0.84 & - & - \\
        & $\textit{CLAP}_{T}$ only & 4.38 & 0.20 & 0.05 & 0.41 & - & 0.24 \\
        & No Conditioning         & 6.70 & 0.27 & 0.03 & 0.25 & - & - \\
    \midrule
    \multirow{5}{*}{\textit{Diff-A-Riff} $_{T=10}^{\text{CFG}=1}$} 
        & $\textit{CLAP}_{A} + \textit{Context}$  & 1.50 & 0.06 & 0.09 & 0.54 & 0.54 & - \\
        & $\textit{CLAP}_{T} + \textit{Context}$  & 6.23 & 0.19 & 0.03 & 0.17 & 0.00  & 0.23\\
        & \textit{Context} only   & 6.59 & 0.25 & 0.03 & 0.24 & 0.00 & - \\
        & $\textit{CLAP}_{A}$ only     & 1.57 & 0.06 & 0.09 & 0.52 & - & - \\
        & $\textit{CLAP}_{T}$ only     & 6.26 & 0.22 & 0.03 & 0.20 & - & 0.22 \\
        & No Conditioning       & 7.67 & 0.28 & 0.03 & 0.20 & - & -\\
  \bottomrule
    \multicolumn{8}{l}{$^{\mathrm{a}}$\(\times{10^{-4}}\), $^b$ obtained from white noise, $^c$ obtained by using a random accompaniment from the dataset}
 \end{tabular}
\end{center}
 \caption{Objective metrics using two configurations, \textit{Diff-A-Riff}$_{T=30}^{\text{CFG}=1.25}$ and \textit{Diff-A-Riff}$_{T=10}^{\text{CFG}=1}$, and different conditional settings (see Sec.~\ref{sec:cond-inference}). We compare against higher bounds obtained from the \emph{real} validation set, and lower bounds obtained from random \textit{noise} or random pairs (\textit{Real}, \textit{Random acc.}). Some cells are empty for APA and CS in the case of context and text-free generation respectively.} 
 \label{tab:obj_metrics}
 \end{small}

\end{table*}

\subsubsection{Listening Tests}
\label{sec:listening_tests}
\noindent \textbf{Subjective Audio Quality (SAQ)}: We perform a Mean Opinion Score (MOS) test to assess audio quality. Participants were presented with 5-second audio segments from real data as well as generations from the baselines and the proposed system. Their task is to rate the audio quality of these segments on a 5-level Likert scale ranging from poor (1) to excellent quality (5). For all items (real data, Diff-A-Riff, and baselines), we compare both complete music pieces as well as solo instruments. 

We generate solo instruments with Diff-A-Riff by conditioning the model on text or audio-derived CLAP embeddings ($\textit{CLAP}_{A\text{ or }T}$) and without input context (for a fair comparison with the baselines, which do not rely on contextual audio inputs). 
Despite the model not being trained for this task, we can also generate \emph{complete} music pieces using \textit{CLAP} and \textit{Context} embeddings, following an iterative approach: First, we create sets of $\textit{CLAP}_{A\text{ or }T}$ embeddings as described in \ref{sec:cond-inference}. Then, from an initially empty context, we generate new tracks from such \textit{CLAP} embeddings, iteratively summing the resulting generation into successive input contexts.

We compare Diff-A-Riff against three state-of-the-art text-to-music baselines: AudioLDM2~\cite{audioldm2}, MusicGen~\cite{musicgen}, and Stable Audio~\cite{stableaudio}.\footnote{We use the open-source ‘AudioLDM2-48kHz’ model. For MusicGen, we use the open-source model ‘MusicGen-large’ operating at 32\,kHz, and for Stable Audio, we use their public API.} For each baseline, we generate 20 5-second excerpts of complete music and solo instruments using text prompts generated by ChatGPT.

 \vspace{0.2cm}
\noindent \textbf{Subjective Audio Prompt Adherence (SAPA)}: We also conduct a subjective assessment of audio-prompt adherence. Participants are provided with a reference 10-second music segment and are asked to rate the compatibility of five distinct accompaniments on a scale from 0 (indicating no adherence) to 100 (perfect adherence), according to harmonic, rhythmic, and overall music style compatibility. 

The reference segments are derived from music pieces within the evaluation set by summing all tracks in a multitrack, excluding one track, which is reserved to serve as the original accompaniment. Each accompaniment is presented mixed with the reference segment, with slight panning applied to the right to aid in distinguishing between them. 
The five accompaniments include the original accompaniment, a randomly selected one from the evaluation set, and three generated by our model under different conditional setups: ($\textit{CLAP}_{A}$ + \textit{Context}), ($\textit{CLAP}_{T}$ + \textit{Context}) and (\textit{Context} only).
To remove a potential bias toward better-quality audio, the original and random segments are encoded and decoded through the CAE.

For both studies, we used the GoListen platform \cite{golisten}. All audio segments are normalized to a loudness of -20 dB LUFS and not cherry-picked. Sample questions are available on the accompanying website.

\section{Results \& Discussion}
\label{sec:results}


\subsection{Objective Evaluation}
\label{sec:obj_eval_res}
\definecolor{mycolor}{RGB}{100, 150, 50} 
\definecolor{customblue}{RGB}{100, 50, 150} 
\definecolor{custom3}{RGB}{150, 50, 0} 
\definecolor{custom4}{RGB}{200, 50, 100} 
\definecolor{custom5}{RGB}{150, 150, 150} 
\definecolor{mycolor}{RGB}{100, 150, 50} 
\definecolor{customblue}{RGB}{100, 50, 150} 
\definecolor{custom6}{RGB}{128, 200, 10} 

\usepgfplotslibrary{fillbetween}
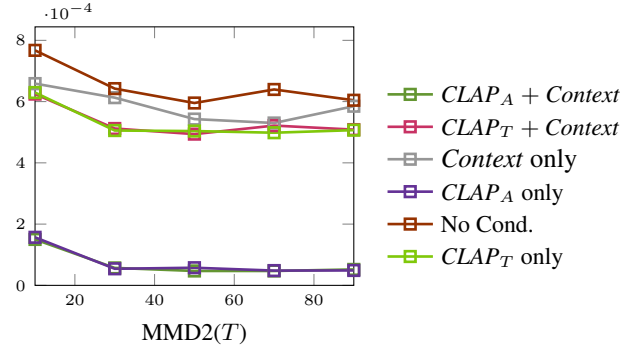
\begin{figure}[t]
\centering
\begin{tikzpicture}
  \begin{groupplot}[
    group style={
      group size=1 by 1,
      horizontal sep=20pt,
    },
    tick label style={font=\tiny},
    xmin=10, xmax=90,
    ymin=0,
    height=5cm
  ]

  \nextgroupplot[
    xlabel={\small{MMD2($T$)}}
  ]
  \addplot [mycolor, mark=square, line width=0.35mm] table [x=X, y=default]{defaultCyran.txt};
  \label{plots:wCLAPaudio}
  \addplot [custom4, mark=square, line width=0.35mm] table [x=X, y=text]{defaultCyran.txt};
  \label{plots:wCLAPtext}
  \addplot [custom5, mark=square, line width=0.35mm] table [x=X, y=no_clap]{defaultCyran.txt};
  \label{plots:woCLAP}
  \addplot [customblue, mark=square, line width=0.35mm] table [x=X, y=no_mix]{defaultCyran.txt};
  \label{plots:woMix}
  \addplot [custom3, mark=square, line width=0.35mm] table [x=X, y=uncond]{defaultCyran.txt};
  \label{plots:uncond}
  \addplot [custom6, mark=square, line width=0.35mm] table [x=X, y=text_only]{defaultCyran.txt};
  \label{plots:text_only}

  \end{groupplot}

    

    \node[right, align=left, inner sep=10pt, anchor=west] at (current bounding box.east) {
    \ref{plots:wCLAPaudio} \small{$\textit{CLAP}_{A} + \textit{Context}$} \quad\\
    \ref{plots:wCLAPtext} \small{$\textit{CLAP}_{T} + \textit{Context}$} \\
    \ref{plots:woCLAP} \textit{Context} only\quad\\
    \ref{plots:woMix} \small{$\textit{CLAP}_{A}$ only} \quad\\
    \ref{plots:uncond} \small{No Cond.}\quad\\
    \ref{plots:text_only} \small{$\textit{CLAP}_{T}$ only}
  };
\end{tikzpicture}

\caption{MMD2 as a function of the number of denoising steps $T$ for various conditional settings (see Sec.~\ref{sec:obj_eval_res}).}
\label{fig:side_by_side_plots}
\end{figure}

Fig.~\ref{fig:side_by_side_plots} shows the MMD2 score of our model as a function of the number of denoising steps, with each line corresponding to a different conditional setting (see Sec.~\ref{sec:cond-inference}), all without classifier-free guidance (CFG).
Note that MMD2 compares the distributions of embeddings of generated and real audio in CLAP's latent space. This means it indicates not only audio quality but also how well the distributions of generated instrument types and timbres match the test data distribution. For this reason, when using audio CLAP conditioning ($\textit{CLAP}_{A}$), the results are considerably better, as we force the timbre distribution to be equal to the test data distribution (by using embeddings of that distribution as conditioning).
However, the improvement of the results when increasing the number of denoising steps can be considered independent of the timbre distribution. 

Based on the results described above we perform a grid search over diffusion steps ($T$) and multi-source classifier-free guidance strength ($CFG_{<source>}$). $T=30$ steps and  $\text{CFG}_\textit{Context} = \text{CFG}_\textit{CLAP} = 1.25$ yields the best results. We denote this configuration as $\textit{Diff-A-Riff}_{T=30}^{\text{CFG}=1.25}$ in Tab.~\ref{tab:obj_metrics}.
In addition, we compare to a specific configuration that achieves real-time performance\footnote{95 seconds of audio in 73 seconds on an AMD EPYC 7502P} on a CPU with acceptable quality using $T=10$ steps and $\text{CFG}=1$, denoted as $\textit{Diff-A-Riff}_{T=10}^{\text{CFG}=1}$. 
For reference, we also include metrics computed on \textit{real} data and a \textit{lower bound}, calculated from white noise in the case of quality metrics or random real pairs for input adherence metrics (APA and CS).
The overall trend suggests that dense conditioning information helps the model in all metrics: audio quality, \textit{coverage} and \textit{density}, as well as APA.
Also, the results could suggest a slight dependency of the model on $\textit{CLAP}_{A}$ embeddings, with metrics close to real data, even in the absence of music context. Only \textit{density} exhibits a minor drop without context conditioning, suggesting that generating from a silent context leads to samples that sometimes fall in low-density regions of the CLAP space. 


The impact of Classifier-free Guidance (CFG) on MMD2 can be estimated by comparing Fig.~\ref{fig:side_by_side_plots}, that displays results without CFG, with Tab.~\ref{tab:obj_metrics}, where CFG was used. In Fig.~\ref{fig:side_by_side_plots}, the metrics for both ($\textit{CLAP}_{A} + \textit{Context}$) and ($\textit{CLAP}_{A}$ only) converge towards an MMD2 of $0.5$, while the corresponding values in Tab.~\ref{tab:obj_metrics} show a reduction to about half this figure. For ($\textit{CLAP}_{T} + \textit{Context}$) and (\emph{$\textit{CLAP}_{T}$ only}), the MMD2 drops from approximately 5 to about 4 with CFG, and for \textit{Context only}, it decreases from around 6 to approximately 5.


Finally, we calculate the Clap Score (CS) for text-conditioned generation ($\textit{CLAP}_{T}$ + $\textit{Context}$, \emph{$\textit{CLAP}_{T}$ only}). We compare $\textit{Diff-A-Riff}_{T=30}^{\text{CFG}=1.25}$ against random pairs of text and real audio (CS=-0.07), suggesting that the model is only somewhat responsive to text prompts (CS=0.25). 

\subsection{Subjective Evaluation}

Table~\ref{tab:saq} presents the outcomes of the Subjective Audio Quality (SAQ) test based on $74$ users who each rated $32$ audio segments, resulting in $2368$ ratings. In this test, we compare results against leading baselines (see Sec.~\ref{sec:listening_tests}) and \emph{real} audio data. Our analysis includes a comparison between these benchmarks and the audio generated by Diff-A-Riff in both mono (ch=1) and pseudo-stereo (ch=2) formats (see Sec.~\ref{sec:cond-inference}). Results show that the pseudo-stereo samples generated by Diff-A-Riff received ratings that are statistically indifferent from \emph{real} audio ratings (p-value=0.79), indicating that participants found the audio quality of the generations indistinguishable from real data. 
This outcome is particularly remarkable given that Diff-A-Riff was not explicitly trained on complete musical pieces nor stereo music generation, but is still competitive to other models. Further, it highlights the influence of stereo imaging on the perceived audio quality.

Tab.~\ref{tab:sapa} shows the results for the Subjective Audio Prompt Adherence (SAPA) listening test based on 35 users, each rating 25 accompaniments. The results include ratings scored by \emph{real} accompaniments, 
\emph{random} accompaniments, 
and the various conditional settings described in Sec.~\ref{sec:cond-inference}.
Following the trend of previous results, the default setting ($\textit{CLAP}_{A} + \textit{Context}$) scores the closest to \textit{real} accompaniments,
suggesting that the model can effectively adapt to the context under this setting. When conditioned on
(\textit{Context} only) 
, Diff-A-Riff is rated worse but still significantly better than \textit{random} accompaniments. Further, for ($\textit{CLAP}_{T} + \textit{Context}$), the accompaniments are rated the lowest. A reason could be that the overall quality is worse because CLAP embeddings of text prompts have not been shown during training. Another problem could be that the randomly chosen text prompt is incompatible with the provided music context (e.g., "A drum machine with electronic textures" with an acoustic blues context), which reduces perceived adherence due to conflicting styles.

Overall, SAPA results are interesting given that APA (see Tab.~\ref{tab:obj_metrics}) suggested rather pessimistic results for ($\textit{CLAP}_{T} + \textit{Context}$)  and (\textit{Context} only). This could potentially be attributed to APA's sensitivity to audio quality and timbre differences between reference and candidate sets.

\begin{table}[t]
 \centering
 \begin{footnotesize}






  \begin{tabular}{cccccc} 

   & SR/Ch & Params  & RTF$^a$ & Solo & Songs \\ 
  \toprule

        \tiny{Real data}   & 44.1/\textbf{2} & - & - & 3.5 ± 0.2  & 3.8 ± 0.2\\
    \midrule
        \tiny{MusicGen}      & 32/1 & 3.3B & 0.4 & 3.1 ± 0.2  & 3.2 ± 0.2 \\
        \tiny{StableAudio}  & 44.1/\textbf{2} & 1B & 11.8 & 2.5 ± 0.2 & 3.0 ± 0.2\\
        \tiny{AudioLDM2}    & \textbf{48}/1  & 712M & 0.4 & 2.6 ± 0.2 & 2.0 ± 0.2\\



    \midrule
        \multirow{2}{*}{\tiny{Diff-A-Riff}}  
            & \textbf{48}/\textbf{2} & \multirow{2}{*}{\textbf{500M}} & \textbf{13.5} (0.57) & \textbf{3.4 ± 0.1} & \textbf{3.8 ± 0.1} \\
             & \textbf{48}/1  & & \textbf{19} (1.3) & 2.8 ± 0.1 & 3.2 ± 0.1 \\
    \bottomrule


    \multicolumn{6}{l}{$^{\mathrm{a}}$\tiny{NVIDIA A100 (CPU : AMD EPYC 7502P)}}\\
 \end{tabular}
    
 \caption{Comparison of Diff-A-Riff to 
 baselines. We include sampling rate in kHz and number of channels (\textit{SR/Ch}), the total number of parameters \textit{Params} (without CLAP
), the Real Time Factor (\textit{RTF}, the ratio of generated time over inference time, for 95 second-long audios) on GPU (and CPU for our model), as well as the SAQ values and 95\% confidence intervals for the subjective audio quality assessment of \textit{Solo} instruments and complete \textit{Songs}.}

 \label{tab:saq}
 \end{footnotesize}
\end{table}

\begin{table}[t]
\centering
 \begin{footnotesize}
 \begin{tabular}{ccc} 

   & Cond. Signal & SAPA \\
  \toprule
        \emph{Real}   & - & 70.1 ± 4.5\\
        \emph{Random}   & - & 12.3 ± 3.0\\
    \midrule
        \multirow{3}{*}{\emph{Diff-A-Riff}}    
            & $\textit{CLAP}_{A} + \textit{Context}$& \textbf{62.4 ± 4.4}\\
            &$\textit{CLAP}_{T} + \textit{Context}$& 37.6 ± 4.2\\
            &\textit{Context} only&  42.3 ± 4.3 \\
        
    \bottomrule
 \end{tabular}
 \caption{Results for SAPA (see Sec.~\ref{sec:listening_tests}). The table includes results of Diff-A-Riff using different conditional settings, with 95\% confidence intervals.}
 \label{tab:sapa}
 \end{footnotesize}
 \vspace{-0.3cm}
\end{table}

\subsection{Control Mechanisms}
In the accompanying website, we show examples of \textit{Diff-A-Riff} generations for different inference settings (see Sec.~\ref{sec:cond-inference}).
We also showcase other controls that naturally emerge from the denoising process, such as \textit{in/out-painting} or the generation of \textit{variations} and \textit{loops}, as well as controls derived from the manipulation of CLAP embeddings, e.g., text-audio \textit{Interpolations}.





\section{Conclusion}

\label{sec:conclusion}
We introduced Diff-A-Riff, a Latent Diffusion Model capable of generating instrumental accompaniments adapted to user-provided musical audio contexts. It can be controlled based on style audio references, text prompts, or both. We also proposed a simple method for producing pseudo-stereo audio.
By exploiting the efficiency of a Consistency Autoencoder, Diff-A-Riff can generate 48\,kHz sample rate pseudo-stereo audio with unprecedented speed and quality. Extensive evaluation experiments showed that our model achieves state-of-the-art audio quality, adapts to various conditional settings, and generates content that adheres to pre-existing musical audio contexts. We believe this work represents a significant step towards AI-assisted music production tools that prioritize artist-centric interactions, enriching the landscape of human-machine music co-creation.


\clearpage

\section{Ethics Statement}
Sony Computer Science Laboratories is committed to exploring the positive applications of AI in music creation. We collaborate with artists to develop innovative technologies that enhance creativity. We uphold strong ethical standards and actively engage with the music community and industry to align our practices with societal values. Our team is mindful of the extensive work that songwriters and recording artists dedicate to their craft. Our technology must respect, protect, and honour this commitment.

Diff-A-Riff supports and enhances human creativity and emphasises the artist's agency by providing various controls for generating and manipulating musical material. By generating a stem at a time, the artist remains responsible for the entire musical arrangement.

Diff-A-Riff has been trained on a dataset that was legally acquired for internal research and development; therefore, neither the data nor the model can be made publicly available. We are doing our best to ensure full legal compliance and address all ethical concerns.


\bibliography{ISMIR2024_template}

\begin{thebibliography}{10}
\providecommand{\url}[1]{#1}
\csname url@samestyle\endcsname
\providecommand{\newblock}{\relax}
\providecommand{\bibinfo}[2]{#2}
\providecommand{\BIBentrySTDinterwordspacing}{\spaceskip=0pt\relax}
\providecommand{\BIBentryALTinterwordstretchfactor}{4}
\providecommand{\BIBentryALTinterwordspacing}{\spaceskip=\fontdimen2\font plus
\BIBentryALTinterwordstretchfactor\fontdimen3\font minus \fontdimen4\font\relax}
\providecommand{\BIBforeignlanguage}[2]{{%
\expandafter\ifx\csname l@#1\endcsname\relax
\typeout{** WARNING: IEEEtran.bst: No hyphenation pattern has been}%
\typeout{** loaded for the language `#1'. Using the pattern for}%
\typeout{** the default language instead.}%
\else
\language=\csname l@#1\endcsname
\fi
#2}}
\providecommand{\BIBdecl}{\relax}
\BIBdecl

\bibitem{makeavideo}
U.~Singer, A.~Polyak, T.~Hayes, X.~Yin, J.~An, S.~Zhang, Q.~Hu, H.~Yang, O.~Ashual, O.~Gafni, D.~Parikh, S.~Gupta, and Y.~Taigman, ``{Make-A-Video}: Text-to-video generation without text-video data,'' in \emph{Proc. of the 11th International Conference on Learning Representations, {ICLR}}, 2023.

\bibitem{dalle3}
J.~Betker, G.~Goh, L.~Jing, TimBrooks, J.~Wang, L.~Li, LongOuyang, JuntangZhuang, JoyceLee, YufeiGuo, WesamManassra, PrafullaDhariwal, CaseyChu, YunxinJiao, and A.~Ramesh, ``Improving image generation with better captions,'' in \emph{CoRR}, 2023.

\bibitem{musiclm}
A.~Agostinelli, T.~I. Denk, Z.~Borsos, J.~H. Engel, M.~Verzetti, A.~Caillon, Q.~Huang, A.~Jansen, A.~Roberts, M.~Tagliasacchi, M.~Sharifi, N.~Zeghidour, and C.~H. Frank, ``{MusicLM: Generating Music From Text},'' in \emph{CoRR}, 2023.

\bibitem{musicgen}
J.~Copet, F.~Kreuk, I.~Gat, T.~Remez, D.~Kant, G.~Synnaeve, Y.~Adi, and A.~D{\'{e}}fossez, ``Simple and controllable music generation,'' in \emph{Advances in Neural Information Processing Systems 36 {NeurIPS}}, 2023.

\bibitem{stableaudio}
Z.~Evans, C.~Carr, J.~Taylor, S.~H. Hawley, and J.~Pons, ``Fast timing-conditioned latent audio diffusion,'' in \emph{CoRR}, 2024.

\bibitem{audioldm2}
H.~Liu, Q.~Tian, Y.~Yuan, X.~Liu, X.~Mei, Q.~Kong, Y.~Wang, W.~Wang, Y.~Wang, and M.~D. Plumbley, ``{AudioLDM 2}: {L}earning holistic audio generation with self-supervised pretraining,'' in \emph{arXiv}, 2023.

\bibitem{clap}
B.~Elizalde, S.~Deshmukh, M.~A. Ismail, and H.~Wang, ``{CLAP} learning audio concepts from natural language supervision,'' in \emph{Proc. of the {IEEE} International Conference on Acoustics, Speech and Signal Processing {ICASSP}}, 2023.

\bibitem{m2l2}
M.~Pasini, S.~Lattner, and G.~Fazekas, ``Music2latent: Consistency autoencoders for latent audio compression,'' in \emph{Proc. of the International Society for Music Information Retrieval (ISMIR)}, 2024.

\bibitem{edm}
T.~Karras, M.~Aittala, T.~Aila, and S.~Laine, ``Elucidating the design space of diffusion-based generative models,'' in \emph{Proc. of the 36th Conference on Neural Information Processing Systems (NeurIPS)}, 2022.

\bibitem{wavenet}
A.~van~den Oord, S.~Dieleman, H.~Zen, K.~Simonyan, O.~Vinyals, A.~Graves, N.~Kalchbrenner, A.~W. Senior, and K.~Kavukcuoglu, ``{WaveNet}: {A} generative model for raw audio,'' in \emph{Proc. of the 9th {ISCA} Speech Synthesis Workshop}, 2016.

\bibitem{samplernn}
S.~Mehri, K.~Kumar, I.~Gulrajani, R.~Kumar, S.~Jain, J.~Sotelo, A.~C. Courville, and Y.~Bengio, ``{SampleRNN}: {A}n unconditional end-to-end neural audio generation model,'' in \emph{Proc. of 5th International Conference on Learning Representations, {ICLR}}, 2017.

\bibitem{gan}
I.~J. Goodfellow, J.~Pouget{-}Abadie \emph{et~al.}, ``Generative adversarial nets,'' in \emph{Advances in Neural Information Processing Systems 27}, Dec. 2014, pp. 2672--2680.

\bibitem{vae}
D.~P. Kingma and M.~Welling, ``Auto-encoding variational bayes,'' in \emph{2nd International Conference on Learning Representations ({ICLR})}, Apr. 2014.

\bibitem{rave}
A.~Caillon and P.~Esling, ``{RAVE:} {A} variational autoencoder for fast and high-quality neural audio synthesis,'' in \emph{CoRR}, 2021.

\bibitem{musika}
M.~Pasini and J.~Schl{\"{u}}ter, ``Musika! {F}ast infinite waveform music generation,'' in \emph{Proc. of the 23rd International Society for Music Information Retrieval Conference, {ISMIR}}, 2022.

\bibitem{drumgan}
J.~Nistal, S.~Lattner, and G.~Richard, ``{DrumGAN:} {S}ynthesis of drum sounds with timbral feature conditioning,'' in \emph{Proc. of the 21st International Society for Music Information Retrieval Conference, {ISMIR}}, 2020.

\bibitem{drumganvst}
J.~Nistal, C.~Aouameur, I.~Velarde, and S.~Lattner, ``{DrumGAN VST:} {A} plugin for drum sound analysis/synthesis with autoencoding generative adversarial networks,'' in \emph{Proc. of International Conference on Machine Learning {ICML}, Workshop on Machine Learning for Audio Synthesis, {MLAS}}, 2022.

\bibitem{crash}
S.~Rouard and G.~Hadjeres, ``{CRASH:} raw audio score-based generative modeling for controllable high-resolution drum sound synthesis,'' in \emph{Proc. of the 22nd International Society for Music Information Retrieval Conference, {ISMIR}}, 2021.

\bibitem{noise2music}
Q.~Huang, D.~S. Park, T.~Wang, T.~I. Denk, A.~Ly, N.~Chen, Z.~Zhang, Z.~Zhang, J.~Yu, C.~H. Frank, J.~H. Engel, Q.~V. Le, W.~Chan, and W.~Han, ``{Noise2Music}: {T}ext-conditioned music generation with diffusion models,'' in \emph{CoRR}, 2023.

\bibitem{edmsound}
G.~Zhu, Y.~Wen, M.~Carbonneau, and Z.~Duan, ``{EDMSound}: {S}pectrogram based diffusion models for efficient and high quality audio synthesis,'' in \emph{{CoRR}}, 2023.

\bibitem{archiSound}
F.~Schneider, ``{ArchiSound}: {A}udio generation with diffusion,'' in \emph{{CoRR}}, 2023.

\bibitem{jukebox}
P.~Dhariwal, H.~Jun, C.~Payne, J.~W. Kim, A.~Radford, and I.~Sutskever, ``Jukebox: {A} generative model for music,'' in \emph{{CoRR}}, 2020.

\bibitem{stemgen}
J.~D. Parker, J.~Spijkervet, K.~Kosta, F.~Yesiler, B.~Kuznetsov, J.~Wang, M.~Avent, J.~Chen, and D.~Le, ``{StemGen}: {A} music generation model that listens,'' in \emph{{CoRR}}, 2023.

\bibitem{bassnet2}
M.~Pasini, M.~Grachten, and S.~Lattner, ``Bass accompaniment generation via latent diffusion,'' in \emph{{IEEE} International Conference on Acoustics, Speech and Signal Processing {ICASSP}}.\hskip 1em plus 0.5em minus 0.4em\relax {IEEE}, 2024.

\bibitem{soundstream}
N.~Zeghidour, A.~Luebs, A.~Omran, J.~Skoglund, and M.~Tagliasacchi, ``{SoundStream}: An end-to-end neural audio codec,'' in \emph{{IEEE} {ACM} Trans. Audio Speech Lang. Process.}, 2022.

\bibitem{encodec}
A.~D{\'{e}}fossez, J.~Copet, G.~Synnaeve, and Y.~Adi, ``High fidelity neural audio compression,'' in \emph{CoRR}, 2022.

\bibitem{vampnet}
H.~F. Garc{\'{\i}}a, P.~Seetharaman, R.~Kumar, and B.~Pardo, ``{VampNet}: Music generation via masked acoustic token modeling,'' in \emph{Proc. of the 24th International Society for Music Information Retrieval Conference, {ISMIR}}, 2023.

\bibitem{magnet}
A.~Ziv, I.~Gat, G.~L. Lan, T.~Remez, F.~Kreuk, A.~D{\'{e}}fossez, J.~Copet, G.~Synnaeve, and Y.~Adi, ``Masked audio generation using a single non-autoregressive transformer,'' in \emph{{CoRR}}, 2024.

\bibitem{mousai}
F.~Schneider, O.~Kamal, Z.~Jin, and B.~Schölkopf, ``Moûsai: Text-to-music generation with long-context latent diffusion,'' in \emph{{CoRR}}, 2023.

\bibitem{jen1}
P.~Li, B.~Chen, Y.~Yao, Y.~Wang, A.~Wang, and A.~Wang, ``{JEN-1:} text-guided universal music generation with omnidirectional,'' in \emph{{CoRR}}, 2023.

\bibitem{ddim}
J.~Song, C.~Meng, and S.~Ermon, ``Denoising diffusion implicit models,'' in \emph{Proc. of the 9th International Conference on Learning Representations, {ICLR}}, 2021.

\bibitem{Jukedrummer}
Y.~Wu, C.~Chiu, and Y.~Yang, ``{Jukedrummer}: Conditional beat-aware audio-domain drum accompaniment generation via transformer {VQ-VAE},'' in \emph{Proc. of the 3rd International Society for Music Information Retrieval Conference, {ISMIR}}, 2022.

\bibitem{mcontrolnet}
S.-L. Wu, C.~Donahue, S.~Watanabe, and N.~J. Bryan, ``Music controlnet: Multiple time-varying controls for music generation,'' in \emph{{CoRR}}, 2023.

\bibitem{musicmagnus}
Y.~Zhang, Y.~Ikemiya, G.~Xia, N.~Murata, M.~A.~M. Ram{\'{\i}}rez, W.~Liao, Y.~Mitsufuji, and S.~Dixon, ``Musicmagus: Zero-shot text-to-music editing via diffusion models,'' \emph{CoRR}, 2024.

\bibitem{manor}
H.~Manor and T.~Michaeli, ``Zero-shot unsupervised and text-based audio editing using {DDPM} inversion,'' \emph{CoRR}, 2024.

\bibitem{ditto}
Z.~Novack, J.~J. McAuley, T.~Berg{-}Kirkpatrick, and N.~J. Bryan, ``{DITTO:} diffusion inference-time t-optimization for music generation,'' \emph{CoRR}, 2024.

\bibitem{cmp}
M.~Levy, B.~D. Giorgi, F.~Weers, A.~Katharopoulos, and T.~Nickson, ``Controllable music production with diffusion models and guidance gradients,'' \emph{CoRR}, 2023.

\bibitem{bassnet}
M.~Grachten, S.~Lattner, and E.~Deruty, ``{BassNet}: A variational gated autoencoder for conditional generation of bass guitar tracks with learned interactive control,'' in \emph{Applied Sciences}, 2020.

\bibitem{singsong}
C.~Donahue, A.~Caillon, A.~Roberts, E.~Manilow, P.~Esling, A.~Agostinelli, M.~Verzetti, I.~Simon, O.~Pietquin, N.~Zeghidour, and J.~H. Engel, ``Singsong: Generating musical accompaniments from singing,'' in \emph{{CoRR}}, 2023.

\bibitem{DrumNet}
S.~Lattner and M.~Grachten, ``High-level control of drum track generation using learned patterns of rhythmic interaction,'' in \emph{2019 {IEEE} Workshop on Applications of Signal Processing to Audio and Acoustics, {WASPAA}}.\hskip 1em plus 0.5em minus 0.4em\relax {IEEE}, 2019.

\bibitem{postolache}
E.~Postolache, G.~Mariani, L.~Cosmo, E.~Benetos, and E.~Rodol{\`{a}}, ``Generalized multi-source inference for text conditioned music diffusion models,'' \emph{CoRR}, 2024.

\bibitem{multisourceddm}
G.~Mariani, I.~Tallini, E.~Postolache, M.~Mancusi, L.~Cosmo, and E.~Rodol{\`{a}}, ``Multi-source diffusion models for simultaneous music generation and separation,'' \emph{CoRR}.

\bibitem{consistency_models}
Y.~Song, P.~Dhariwal, M.~Chen, and I.~Sutskever, ``Consistency models,'' in \emph{International Conference on Machine Learning, {ICML}}, 2023.

\bibitem{improved_consistency_models}
Y.~Song and P.~Dhariwal, ``Improved techniques for training consistency models,'' \emph{arXiv preprint arXiv:2310.14189}, 2023.

\bibitem{ddpm}
J.~Ho, A.~Jain, and P.~Abbeel, ``Denoising diffusion probabilistic models,'' in \emph{Advances in Neural Information Processing Systems, {NeurIPS}}, 2020.

\bibitem{diffusion_original}
J.~Sohl{-}Dickstein, E.~A. Weiss \emph{et~al.}, ``Deep unsupervised learning using nonequilibrium thermodynamics,'' in \emph{Proceedings of the 32nd International Conference on Machine Learning, {ICML}}, 2015.

\bibitem{NCSN++}
T.~Karras, M.~Aittala, T.~Aila, and S.~Laine, ``Score-based generative modeling through stochastic differential equations,'' in \emph{Proc. of the International Conference on Learning Representations (ICLR)}, 2021.

\bibitem{adamw}
I.~Loshchilov and F.~Hutter, ``Decoupled weight decay regularization,'' in \emph{7th International Conference on Learning Representations, {ICLR}}, 2019.

\bibitem{cfg}
J.~Ho and T.~Salimans, ``Classifier-free diffusion guidance,'' \emph{arXiv preprint arXiv:2207.12598}, 2022.

\bibitem{kid}
M.~Binkowski, D.~J. Sutherland, M.~Arbel, and A.~Gretton, ``Demystifying {MMD} gans,'' in \emph{Proc. of the 6th International Conference on Learning Representations, {ICLR}}, 2018.

\bibitem{fad}
K.~Kilgour, M.~Zuluaga, D.~Roblek, and M.~Sharifi, ``Fr{\'{e}}chet audio distance: {A} reference-free metric for evaluating music enhancement algorithms,'' in \emph{Proc. of the 20th Conference of the International Speech Communication Association (InterSpeech)}, 2019.

\bibitem{den_cov}
M.~F. Naeem, S.~J. Oh, Y.~Uh, Y.~Choi, and J.~Yoo, ``Reliable fidelity and diversity metrics for generative models,'' in \emph{Proc. of the 37th International Conference on Machine Learning, {ICML}}, 2020.

\bibitem{Make-An-Audio}
R.~Huang, J.~Huang, D.~Yang, Y.~Ren, L.~Liu, M.~Li, Z.~Ye, J.~Liu, X.~Yin, and Z.~Zhao, ``Make-an-audio: Text-to-audio generation with prompt-enhanced diffusion models,'' in \emph{Proc. of the International Conference on Machine Learning, {ICML}}, 2023.

\bibitem{apa}
M.~Grachten and J.~Nistal, ``{Audio Prompt Adherence}: A measure for evaluating musical accompaniment systems,'' in \emph{{CoRR}}, 2024.

\bibitem{golisten}
D.~Barry, Q.~Zhang, P.~W. Sun, and A.~Hines, ``Go listen: An end-to-end online listening test platform.'' \emph{Journal of Open Research Software}, vol.~9, no.~1, 2021.

\end{thebibliography}
\end{document}